# Weak Coupling Regime of the Landau–Zener Transition for Association of an Atomic Bose–Einstein Condensate[1]

N. Sahakyan[a], H. Azizbekyan[a,b], H. Ishkhanyan[b], R. Sokhoyan[a], and A. Ishkhanyan[a]

[a] *Institute for Physical Research NAS of Armenia, 0203 Ashtarak-2, Armenia*
[b] *Moscow Institute of Physics and Technology, Dolgoprudny, 141700 Russia*
e-mail: aishkhanyan@gmail.com

**Abstract**—In the framework of a basic semiclassical time-dependent nonlinear two-state problem, we study the weak coupling limit of the nonlinear Landau–Zener transition at coherent photo- and magneto-association of an atomic Bose–Einstein condensate. Using an exact third-order nonlinear differential equation for the molecular state probability, we develop a variational approach which enables us to construct an accurate analytic approximation describing time dynamics of the coupled atom-molecular system for the case of weak coupling. The approximation is written in terms of the solution to an auxiliary linear Landau–Zener problem with some effective Landau–Zener parameter. The dependence of this effective parameter on the input Landau–Zener parameter is found to be unexpected: as the generic Landau–Zener parameter increases, the effective Landau–Zener parameter first monotonically increases (starting from zero), reaches its maximal value and then monotonically decreases again reaching zero at some point. The constructed approximation quantitatively well describes many characteristics of the time dynamics of the system, in particular, it provides a highly accurate formula for the final transition probability to the molecular state. The present result for the final transition probability improves the accuracy of the previous approximation by Ishkhanyan et al. [Phys. Rev. A **69**, 043612 (2004); J. Phys. A **38**, 3505 (2005)] by order of magnitude.

**DOI:** 10.1134/S1054660X10010184

## 1. INTRODUCTION

Ultracold quantum gases have become a rapidly developing research domain due to recent experimental and theoretical achievements (see, e.g., the recent reviews [1–3]). One of the remarkable research directions in this field is *coherent* molecule formation in atomic quantum gases via application of associating optical or magnetic fields (such processes are referred to as "superchemistry" [4]) which under certain experimental conditions [5, 6] can be described by a basic mean-field time-dependent two-level problem that is defined by the following set of coupled nonlinear equations [7]:

$$i\frac{da_1}{dt} = U(t)e^{-i\delta(t)}\bar{a}_1 a_2,$$
$$i\frac{da_2}{dt} = \frac{U(t)}{2}e^{i\delta(t)}a_1 a_1, \quad (1)$$

where $t$ is time, $a_1$ and $a_2$ are the atomic and molecular state probability amplitudes, respectively, $\bar{a}_1$ denotes the complex conjugate of $a_1$, the real function $U(t)$ is referred to as the Rabi frequency of the associating field, and the real function $\delta(t)$ is the integral of the associated frequency detuning.

The field configuration $\{U(t), \delta(t)\}$ we discuss in the present paper is the celebrated Landau–Zener (LZ) model [8, 9] (see also [10, 11]) for which the detuning is assumed to be a linear function of time: $\delta_t(t) = 2\delta_0 t$, and the Rabi frequency is supposed to be constant: $U(t) = U_0 = $ const. This is a well appreciated model that serves as a prototype for all the level crossing models [12, 13]. We treat the basic case when the evolution of the system starts from the pure atomic state, so that the initial conditions are

$$p(-\infty) = 0, \quad p_t(-\infty) = 0, \quad p_{tt}(-\infty) = 0, \quad (2)$$

where $p = |a_2|^2$ is the molecular state probability. We examine the weak coupling limit of the association process, which corresponds to the weak associating fields with the Rabi frequency being $U_0 < 1$. This limit has previously been discussed by several authors both in the context of photoassociation and Feshbach resonance (see, e.g., [14–18]) and an accurate formula for the final transition probability to the molecular state at $t \longrightarrow +\infty$ has been proposed [14, 15]. However, the whole time dynamics of the system has not been treated in detail.

In the present paper we suggest a rigorous analysis of the time evolution of the system. Using a variational approach we construct an accurate analytic approximation applicable in the whole time domain. This approximation is written in terms of an auxiliary linear

LZ solution for an effective LZ parameter. The dependence of the latter parameter on the conventional LZ parameter is rather unexpected: as the generic LZ parameter increases, the effective LZ parameter first monotonically increases (starting from zero), reaches the maximum and then monotonically decreases again reaching zero at some point. The constructed solution correctly describes many characteristics of the time evolution of the molecular state probability for the whole reasonable range of variation of the LZ parameter. The approximation accurately describes the behavior of the system before and during the resonance crossing. Furthermore, the frequency of the atom-molecule oscillations that start soon after crossing the resonance is matched with high accuracy. The solution quantitatively correctly describes the oscillation at large time. However, in the time region covering several first oscillation periods deviation of the presented solution from the numerical result is observed. In this region, the amplitude of the oscillations is overestimated by the analytic solution. This deviation increases as the LZ parameter approaches unity. The final molecular probability at $t \rightarrow +\infty$ is also determined with high accuracy. The derived result improves the accuracy of the previous approximation for the final transition probability [14, 15] by order of magnitude.

## 2. EQUATION FOR THE MOLECULAR STATE PROBABILITY

System (1) describes a lossless process, i.e., preserves the total number of particles, which we choose as normalized to unity:

$$|a_1|^2 + 2|a_2|^2 = \text{const} = 1. \quad (3)$$

After eliminating one of the dependent variables from system (1) we obtain the following nonlinear equations of the second order for the atomic and molecular states' probability amplitudes, respectively:

$$a_{1tt} + \left(i\delta_t - \frac{U_t}{U}\right)a_{1t} - \frac{U^2}{2}(1 - 2|a_1|^2)a_1 = 0, \quad (4)$$

$$a_{2tt} + \left(-i\delta_t - \frac{U_t}{U}\right)a_{2t} + U^2(1 - 2|a_2|^2)a_2 = 0. \quad (5)$$

Hence, instead of the two coupled first-order equations (1) one may work with one second-order equation [either (4) or (5)]. However, it turns out that it is more convenient to deal with one equation, written for the molecular state probability $|a_2|^2$ [19, 20]. It can be shown that all other involved variables are then expressed in terms of this quantity.

It can be shown by direct differentiation that the transition probability $p = |a_2|^2$ satisfies the following relations:

$$p_t = a_{2t}^*a_2 + a_2^*a_{2t} = -i\frac{U}{2}(a_1^2 a_2^{*} e^{i\delta(t)} - a_1^{*2}a_2 e^{-i\delta(t)}), \quad (6)$$

$$p_{tt} = \frac{U_t}{U}p_t + \frac{U^2}{2}(1 - 8p + 12p^2) \\ + \frac{U}{2}\delta_t(a_1^2 a_2^{*} e^{i\delta(t)} + a_1^{*2}a_2 e^{-i\delta(t)}). \quad (7)$$

Furthermore, it can be checked by straightforward differentiation that the function

$$Z = a_1^2 a_2^{*} e^{i\delta(t)} + a_1^{*2}a_2 e^{-i\delta(t)} \quad (8)$$

satisfies the relation

$$Z_t = -\delta_t \frac{2p_t}{U}. \quad (9)$$

Then, differentiation of equation (7) followed by some algebra yields the following nonlinear ordinary differential equation of the third order for the molecular state probability [19]:

$$p_{ttt} - \left(\frac{\delta_{tt}}{\delta_t} + 2\frac{U_t}{U}\right)p_{tt} \\ + \left[\delta_t^2 + 4U^2(1 - 3p) - \left(\frac{U_t}{U}\right)_t + \frac{U_t}{U}\left(\frac{\delta_{tt}}{\delta_t} + \frac{U_t}{U}\right)\right]p_t \quad (10) \\ + \frac{U^2}{2}\left(\frac{\delta_{tt}}{\delta_t} - \frac{U_t}{U}\right)(1 - 8p + 12p^2) = 0.$$

It is worth stressing that the normalization condition (3) is incorporated in this equation.

The derived equation is considerably simplified for the models with constant field amplitude: $U(t) = U_0$. In this case we have the equation

$$p_{ttt} - \frac{\delta_{tt}}{\delta_t}p_{tt} + [\delta_t^2 + 4U_0^2(1 - 3p)]p_t \\ + \frac{U_0^2}{2}\frac{\delta_{tt}}{\delta_t}(1 - 8p + 12p^2) = 0 \quad (11)$$

which can be rewritten in the following factorized form [20]:

$$\left(\frac{d}{dt} - \frac{\delta_{tt}}{\delta_t}\right)\left(p_{tt} - \frac{U_0^2}{2}(1 - 8p + 12p^2)\right) + \delta_t^2 p_t = 0. \quad (12)$$

This equation serves as a starting point for the development presented below. For the LZ model, Eq. (12) is written as

$$\left(\frac{d}{dt} - \frac{1}{t}\right)\left(p_{tt} - \frac{\lambda}{2}(1 - 8p + 12p^2)\right) + 4t^2 p_t = 0, \quad (13)$$

where we have passed to the dimensionless time, $t \longrightarrow t/\sqrt{\delta_0}$, and have introduced the conventional LZ parameter

$$\lambda = \frac{U_0^2}{\delta_0}. \quad (14)$$

The *linear* counterpart of the nonlinear system (1) reads

$$i\frac{da_{1L}}{dt} = U_0 e^{-i\delta(t)} a_{2L}, \quad i\frac{da_{2L}}{dt} = U_0 e^{i\delta(t)} a_{1L}. \quad (15)$$

Accordingly, for the second state probability $p_L = |a_{2L}|^2$ of the linear problem we have the equation [compare with Eq. (13)]

$$\left(\frac{d}{dt} - \frac{1}{t}\right)\left(p_{Ltt} - \frac{\lambda}{2}(4 - 8p_L)\right) + 4t^2 p_{Lt} = 0. \quad (16)$$

Note that when deriving Eq. (16) the normalization constraint $|a_{1L}|^2 + |a_{2L}|^2 = 1$ has been taken into account. This linear differential equation is exactly solvable, and we denote as $p_{LZ}$ the solution which satisfies the initial conditions: $p_L(-\infty) = 0$, $p_{Lt}(-\infty) = 0$, $p_{Ltt}(-\infty) = 0$. This solution can be written in terms of the confluent hypergeometric functions [15]:

$$p_{LZ}(t) = |a_{2LZ}(t)|^2 = |C_{01} F_1 + C_{02} F_2|^2 \quad (17)$$

with

$$C_{01} = \sqrt{\lambda e^{-\pi\lambda/4} \cosh(\pi\lambda/4)} \frac{i}{2} \frac{\Gamma(1/2 - i\lambda/4)}{\Gamma(1 - i\lambda/4)},$$
$$C_{02} = \sqrt{\lambda e^{-\pi\lambda/4} \cosh(\pi\lambda/4)} \sqrt{i\delta_0}, \quad (18)$$

and

$$F_1 = {}_1F_1(i\lambda/4; 1/2; it^2),$$
$$F_2 = t \, {}_1F_1(1/2 + i\lambda/4; 3/2; it^2), \quad (19)$$

where $\Gamma$ is the Euler gamma-function [21] and ${}_1F_1$ is the Kummer confluent hypergeometric function [21]. The limits of $p_{LZ}$ for $t \longrightarrow 0$ and $t \longrightarrow +\infty$ are written as

$$p_{LZ}(0) = \frac{1 - e^{-\pi\lambda/2}}{2}, \quad p_{LZ}(+\infty) = 1 - e^{-\pi\lambda}. \quad (20)$$

## 3. MATHEMATICAL TREATMENT AND RESULTS

In what follows we try to construct an approximate solution of the equation for the molecular state probability (13) in the *weak interaction* limit, $\lambda < 1$, using the solution $p_{LZ}$ to the linear equation (16). To do this, we consider the linear equation (16) with an auxiliary parameter $\lambda_1$:

$$\left(\frac{d}{dt} - \frac{1}{t}\right)\left(p_{LZtt} - \frac{\lambda_1}{2}(4 - 8p_{LZ})\right) + 4t^2 p_{LZt} = 0, \quad (21)$$

and try to approximate the solution to the exact equation (13) as follows:

$$p_0 = C_1 p_{LZ}(\lambda_1, t). \quad (22)$$

As it is seen, apart from a simple pre-factor $C_1$, we have here introduced an *effective* LZ parameter $\lambda_1$. After substituting this expression into the left-hand side of equation (13) and taking into account that function $p_{LZ}(\lambda_1, t)$ satisfies equation (21), we get some remainder:

$$R = \left(\frac{d}{dt} - \frac{1}{t}\right) r(t), \quad (23)$$

where $r(t)$ is the notation for

$$r(t) = C_1 \frac{\lambda_1}{2}(4 - 8p_{LZ}(\lambda_1))$$
$$- \frac{\lambda}{2}(1 - 8C_1 p_{LZ}(\lambda_1) + 12(C_1 p_{LZ}(\lambda_1))^2). \quad (24)$$

It is obvious that if the remainder $R$ is identically zero, then the function $p_0$ given by equation (22) is the exact solution to equation (13). Hence, it is intuitively understood that a way to proceed is to try to minimize $R$ by means of appropriate choice of $\lambda_1$ and $C_1$.

To do this, we first note that, since the function $p_{LZ}(\lambda_1, t)$ is bounded everywhere, the function $R$ is bounded almost everywhere. The exception is the resonance crossing point $t = 0$, where, due to the term $1/t$ of the operator $(d/dt - 1/t)$, in general, $R$ diverges. Therefore, as a first step, we eliminate this divergence, i.e., we require $\lambda_1$ and $C_1$ to satisfy the equation $r(0) = 0$. Explicitly, this equation is written as

$$C_1 \frac{\lambda_1}{2}(4 - 8p_{LZ}(\lambda_1, 0))$$
$$- \frac{\lambda}{2}(1 - 8C_1 p_{LZ}(\lambda_1, 0) + 12 C_1^2 p_{LZ}(\lambda_1, 0)^2) = 0. \quad (25)$$

To find appropriate values of parameters $\lambda_1$ and $C_1$ we need to introduce one more equation. Of course, in order to construct an approximation as simple as possible, one may first try to avoid variation of both auxiliary parameters and attempt to get a simpler, one-parametric approximation instead. A natural choice is then to fix $\lambda_1 = \lambda$ and vary $C_1$ alone.

Equation (25) then immediately yields:

$$C_1 = \frac{1 - \sqrt{1 - 3P_{LZ}(0)^2}}{6 P_{LZ}(0)^2} \quad (26)$$

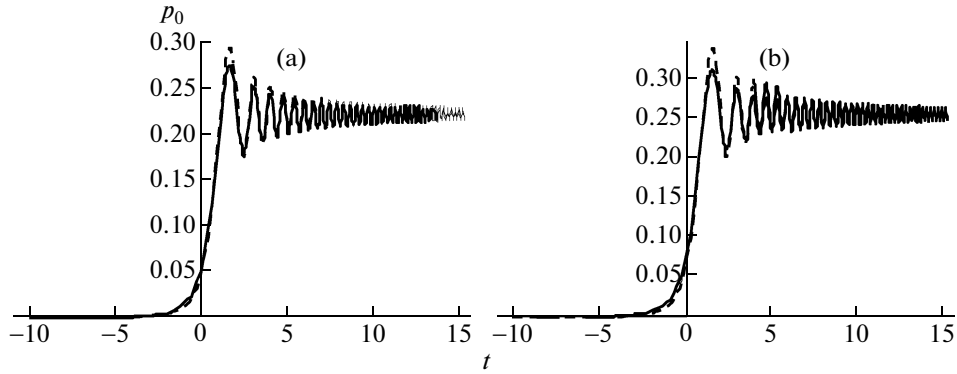

**Fig. 1.** Comparison of the solution (22) (dotted lines) with the numerical solution to equation (13) (solid lines) at (a) $\lambda = 0.25$ and (b) 0.64.

with $P_{LZ}(0)$ given by Eq. (20). Numerical examination reveals that this approximation works only for very small $\lambda$, of the order of $10^{-2}$. For $\lambda = 0.25$ the error is about 5% and for $\lambda = 1$ it becomes 20%. This result shows that we have to consider two-parametric fit, hence, we do need a second equation to determine the parameters $\lambda_1$ and $C_1$.

For derivation of the second equation we write the (exact) solution to the initial equation (13) as

$$p = p_0 + u \qquad (27)$$

and examine the exact equation for the correction term $u$. This equation is written as

$$\left(\frac{d}{dt} - \frac{1}{t}\right)\left(u_{tt} - \frac{\lambda}{2}(-8u + 24p_0 u + 12u^2) + r\right) \qquad (28)$$
$$+ 4t^2 u_t = 0.$$

Now, considering the correction $u$ as sufficiently small (since the zero-order approximation $p_0$ is supposed to be sufficiently close to the exact solution) one arrives at the needed second equation:

$$\left(C_1 \frac{\lambda_1}{2}(4 - 8p_{LZ}(\lambda_1, +\infty))\right.$$
$$\left. - \frac{\lambda}{2}(1 - 8C_1 p_{LZ}(\lambda_1, +\infty) + 12C_1^2 p_{LZ}(\lambda_1, +\infty)^2)\right) \qquad (29)$$
$$+ \left(2C_1\lambda_1 - \frac{\lambda}{2}\right) - \frac{\lambda}{2}p_{LZ}^2(\lambda_1, +\infty) = 0.$$

The details of derivation of this equation are presented in Appendix.

Thus, the values of parameters $\lambda_1$ and $C_1$ for which the function (22) approximates the exact solution to equation (13) are determined by equations (25) and (29). One may numerically solve these equations and further compare the proposed approximation (22) with the numerical solution to the exact equation (13).

The comparison is shown on Fig. 1. It is seen that the coincidence is quite good—some deviation is observed only for the several first oscillations occurring after resonance-crossing and we see that the deviation becomes visible for relatively large $\lambda$: $0.3 \le \lambda \le 1.0$. In the meanwhile, several important characteristics of the process, such as the final transition probability (at $t \longrightarrow +\infty$) and the frequency of oscillations are determined with high accuracy. As is seen, it is the amplitude of oscillations that displays significant deviations from the numerical result. However, it can be shown that there exists a modification of the applied variational method which is potent to provide essential improvement of the result. We will discuss this development in a further paper.

The system of equations (25) and (29) for determination of optimal values of the parameters $\lambda_1$ and $C_1$ can be solved by approximate methods. In order to do this, we first eliminate $C_1$ from the system. Next, we show that with the increase of $\lambda$ the function $\lambda_1(\lambda)$ first monotonically increases starting from $\lambda_1(0) = 0$ and further monotonically decreases to zero at $\lambda = \sqrt{2}$. The function reaches its maximal value at $\lambda \approx 0.454$. The corresponding auxiliary parameter $\lambda_1$ then adopts the value $\lambda_{1\max} \approx 0.124$. A sufficiently good approximation for the function $\lambda_1(\lambda)$ is given by the following formula:

$$\lambda_1 = \lambda\left(1 - \frac{\lambda}{\sqrt{2}}\right)\frac{1 + \lambda/\pi}{1 + 4\lambda}. \qquad (30)$$

Comparison of the derived formula with the numerical result is shown in Fig. 2a. It is seen that the approximation for the interval $\lambda \in [0, 1]$ (i.e., the whole weak interaction limit) is rather good. Furthermore, we apply equation (30) to calculate the function $C_1(\lambda)$ using equation (25). Comparison of the resultant approximation with the numerical result for $C_1(\lambda)$ is shown in Fig. 2b. As it is seen, the graphs are practically indistinguishable.

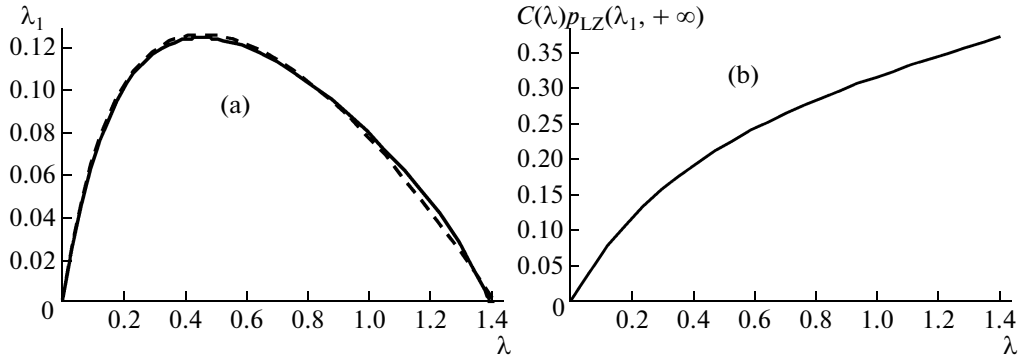

**Fig. 2.** Functions $\lambda_1(\lambda)$ [dotted line-formula (30)] and $C_1(\lambda)p_{LZ}(\lambda_1, +\infty)$.

An analytic expression for $C_1$ as a function of $\lambda$ can be constructed starting from equation (25) and applying, for example, successive iterations. Indeed, rewrite equation (25) in the following form

$$C_1 = \frac{\lambda/4}{\lambda_1 + 2(\lambda - \lambda_1)p_{LZ}(\lambda_1, 0) - 3\lambda C_1 p_{LZ}(\lambda_1, 0)^2}. \quad (31)$$

Now, as a zero-order approximation we put in the right-hand side of this equation $C_1^{(0)} = 0$.

As a result, we get an expression that is already a good approximation. By applying further iterations, we conclude that an accurate approximation for all $\lambda \leq 1$ is given by the formula

$$C_1 = \frac{\lambda/4}{\lambda_1 + 1.75(\lambda - \lambda_1)p_{LZ}(\lambda_1, 0)}. \quad (32)$$

The graph of this function is visually indistinguishable from the numerical solution. The substitution of $\lambda_1(\lambda)$ from Eq. (30) into this formula shows that function $C_1(\lambda)$, in contrast to the function $\lambda_1(\lambda)$, monotonically increases on the segment $\lambda \in [0, 1]$.

The final transition probability to the molecular state at $t \longrightarrow +\infty$ for the weak interaction limit, according to Eq. (22), is $p(+\infty) = C_1(\lambda)p_{LZ}(\lambda_1, +\infty)$. It is namely this function that is shown in Fig. 2b. Note finally that the direct comparison shows that on the whole segment $\lambda \in [0, 1]$ expression (32) produces a final transition probability that accurately matches the formula derived in [14, 15].

## 4. SUMMARY

Thus, we have discussed the coherent photo- and magneto-association of cold atoms with formation of cold diatomic molecules, in the framework of a basic semiclassical nonlinear two-state problem describing a zero-dimensional Bose–Einstein condensate in the mean field approximation. Using an exact third-order nonlinear differential equation for the molecular state probability, we have developed an effective variational method for constructing the approximate solution to the problem in the limit of weak coupling.

Discussing the LZ problem for which the Rabi frequency is constant and the detuning linearly in time crosses the resonance, we have proposed a basic functional form of the appropriate approximation. The approximation is written in the form of a modification of the solution to the corresponding linear LZ problem with two scaling parameters involved. One of these parameters is a simple pre-factor while another one serves as an effective LZ parameter. We have shown that by an appropriate choice of these parameters the proposed solution well describes the main characteristics of the exact solution to the problem in the whole time domain.

We have constructed relevant equations for determination of the introduced auxiliary parameters. The approach for the derivation of these equations is as follows. The proposed zero-order initial approximation generates an inhomogeneous term (which is referred to as remainder) in the exact equation for the next approximation. The further idea was, first, to suppress the divergence of this remainder observed in the resonance-crossing point. Furthermore, second, we minimized the correction term by suppressing, as much as possible, the influence of the remainder on the forming of this next approximation term.

We have derived an accurate analytic approximation to the solution of the formulated equations for the scaling parameters. As expected, the pre-factor $C_1$ is a slowly varying, monotonically increasing function of the LZ parameter. In the meantime, the dependence of the effective LZ parameter $\lambda_1$ on the input LZ parameter $\lambda$ turns to be non-trivial. As $\lambda$ increases, the function $\lambda_1(\lambda)$ first monotonically increases ($\lambda_1 \approx \lambda$) starting from $\lambda_1(0) = 0$, reaches its maximal value, $\lambda_{1\max} \approx 0.124$, at $\lambda \approx 0.454$ and further monotonically decreases reaching zero at $\lambda = \sqrt{2}$.

Using the constructed approximation, we have examined the weak coupling limit of the association process, which corresponds to the weak associating fields: $\lambda \leq 1$. This limit has previously been discussed by several authors. In particular, in [14, 15] a formula for the final transition probability at $t \longrightarrow +\infty$ has been derived. In the present paper we go beyond the previous studies and treat the whole time dynamics of the system. The results are inspiring. The constructed solution qualitatively correctly describes the main characteristics of the evolution of the molecular state probability as a function of time for the supposed range of variation of the LZ parameter: $\lambda \in [0, 1]$. Furthermore, the behavior of the system before and during the resonance crossing and the frequency of the oscillations of the transition probability that start soon after crossing the resonance, as well as the final transition probability to the molecular state are determined with high accuracy. In particular, the present result improves the accuracy of the previous approximation for the final transition probability [14, 15] by order of magnitude. Only the amplitude of the oscillations displays visible deviations from that of the numerical solution. The deviation is mostly pronounced for several first oscillation periods and becomes rather significant when the LZ parameter approaches unity. We hope to address this point in a separate publication.

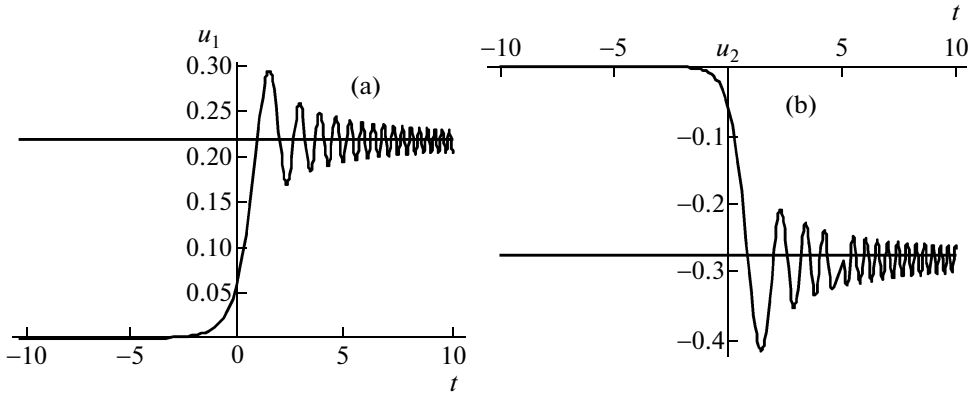

**Fig. 3.** Functions $u_1(t)$ and $u_2(t)$ for $\lambda = 0.0625$.

## APPENDIX

Consider equation (27) obeyed by the correction term $u$:

$$\left(\frac{d}{dt} - \frac{1}{t}\right)\left(u_{tt} - \frac{\lambda}{2}(-8u + 24p_0 u + 12u^2) + r\right) \quad (A1)$$
$$+ 4t^2 u_t = 0.$$

The initial conditions for $u$ are

$$u(-\infty) = 0, \quad u_t(-\infty) = 0, \quad u_{tt}(-\infty) = 0. \quad (A2)$$

Now, examining Eq. (24), we see that since the solution $p_{LZ}$ to the linear LZ problem is a step-wise func-

tion, the remainder $r$ can be presented as a sum

$$r = r_1 + r_2 + r_3, \quad (A3)$$

where $r_1 = r(-\infty) = \text{const}$, $r_2 = r_{20}(1 + \tanh(t))/2$ with $r_{20} = r(+\infty) - r(-\infty) = \text{const}$. It is then understood that the last term, $r_3$, is a relatively small, oscillating quantity. Therefore, supposedly, this term can be neglected.

Furthermore, if the approximate solution $p_0$ is supposed to be good enough so that $|u| \ll p_0 \leq 1/2 \Rightarrow |u| \ll 0.1 \Rightarrow u^2 \ll |u|$, then we can linearize equation (27) by neglecting the quadratic term $12u^2$. Then, since the remaining equation is linear, we may expect that $u(r) \approx u_1(r_1) + u_2(r_2) + u_3(r_3)$, where $u_1$, $u_2$, and $u_3$ are the solutions to the linearized equations written for the remainders $r_1$, $r_2$, and $r_3$, respectively, satisfying the same initial conditions (A2). We have mentioned above that $r_3$ is expected to be small as compared to $r_1$ and $r_2$. Moreover, since it is also an oscillating function, then the solution $u_3(r_3)$ should be small as compared to $u_1(r_1)$ and $u_2(r_2)$. Numerical experiments confirm this assumption. Therefore, we neglect the term $u_3(r_3)$.

The auxiliary linear differential equations for $u_1$ and $u_2$ are

$$\left(\frac{d}{dt} - \frac{1}{t}\right)\left(u_{1tt} - \frac{\lambda}{2}(-8u_1 + 24p_0 u_1) + r_1\right) \quad (A4)$$
$$+ 4t^2 u_{1t} = 0,$$

$$\left(\frac{d}{dt} - \frac{1}{t}\right)\left(u_{2tt} - \frac{\lambda}{2}(-8u_2 + 24p_0 u_2) + r_2\right) \quad (A5)$$
$$+ 4t^2 u_{2t} = 0.$$

In the weak interaction limit the probability $p$ of molecule formation is small, hence, the approximate solution $p_0$ is also relatively small. This means that we can neglect the term $24p_0 u_1$ in Eq. (A4), and, as a result, we will get a LZ-type equation for $u_1$ [compare with equation (21)]:

$$\left(\frac{d}{dt} - \frac{1}{t}\right)\left(u_{1tt} - \frac{\lambda}{2}\left(-\frac{2r_1}{\lambda} - 8u_1\right)\right) + 4t^2 u_{1t} = 0. \quad (A6)$$

The exact solution to this equation is $u_1 = -r_1 p_{LZ}(\lambda)/2\lambda$, therefore $u_1(+\infty) \approx -r_1/(2\lambda) p_{LZ}(\lambda, +\infty)$.

It is also possible to construct a good approximation for the solution to equation (A5) for the function $u_2$. As a result of the construction (using the method "from the inverse") we arrive at the approximation

$$u_2(+\infty) \approx -r_{20}/(4\lambda) p_{LZ}(\lambda, +\infty). \quad (A7)$$

The behavior of functions $u_1(t)$ and $u_2(t)$ is shown in Fig. 3. As it is seen, the correction $u \approx u_1 + u_2$ will be significantly suppressed and, hence, $u$ will be essentially suppressed if we require canceling of $u_1(+\infty)$ and $u_2(+\infty)$:

$$u_1(+\infty) + u_2(+\infty) \approx \lim_{t \to +\infty} u \approx 0. \quad (A8)$$

This condition is equivalent to the relation $r_1/(2\lambda) + r_{20}/(4\lambda) \approx 0$. The latter means that the following equation should hold:

$$r(+\infty) + r(-\infty) \approx 0. \quad (A9)$$

More detailed examination of the roles of $r_1$, $r_2$, and $r_3$ in formation of the correction $u$ shows that for further suppression of the latter we should add in the derived equation a small correction that is of the order of $\lambda^3$. Finally, we arrive at the following relation:

$$r(+\infty) + r(-\infty) \approx \frac{\lambda}{2} p_{LZ}^2(+\infty). \quad (A10)$$

Explicitly, this equation is written as

$$\left(C_1 \frac{\lambda_1}{2}(4 - 8 p_{LZ}(\lambda_1, +\infty))\right.$$
$$\left. - \frac{\lambda}{2}(1 - 8C_1 p_{LZ}(\lambda_1, +\infty) + 12(C_1 p_{LZ}(\lambda_1, +\infty))^2)\right) \quad (A11)$$
$$+ \left(2C_1\lambda_1 - \frac{\lambda}{2}\right) - \frac{\lambda}{2} p_{LZ}^2(\lambda_1, +\infty) = 0.$$

The derived equation is the second equation required for determination of the auxiliary parameters $\lambda_1$ and $C_1$.

ACKNOWLEDGMENTS

This work was supported by the Armenian National Science and Education Fund (ANSEF grant no. 2009-PS-1692) and the International Science and Technology Center (ISTC grant no. A-1241). This research has been conducted in the scope of the International Associated Laboratory IRMAS. R. Sokhoyan and H. Azizbekyan acknowledge the French Embassy in Yerevan for the grants nos. 2006-4638 and 2007-3849 (Boursiers du Gouvernement Français).